\documentclass[runningheads,orivec]{llncs}
\usepackage{amsmath}
\usepackage{amssymb}
\usepackage{graphicx}
\usepackage{xparse}
\usepackage{hyperref}
\usepackage{xcolor}
\usepackage{dsfont}
\usepackage{pifont}
\usepackage{multirow}
\usepackage[T1]{fontenc}
\usepackage{bbding}

\def\orcidID#1{\href{http://orcid.org/#1}{\raisebox{-1.25pt}{\includegraphics{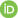}}}}

\renewcommand{\Alph}[0]{\Sigma}
\newcommand{\VarS}[0]{\mathcal{V}_S}
\newcommand{\VarC}[0]{\mathcal{V}_C}
\newcommand{\Tokens}[0]{\mathcal{K}}
\newcommand{\Term}[0]{\mathcal{T}}
\newcommand{\Int}[0]{\mathcal{M}}

\newcommand{\STSubStr}[1]{2^{#1}}

\newcommand{\Len}[1]{|#1|}
\newcommand{\SLen}[1]{||#1||}
\newcommand{\SDec}[1]{\operatorname{pre}(#1)}

\newcommand{\Parikh}[2]{\alpha_{#1}(#2)}
\newcommand{\ParikhBoth}[2]{\alpha^{\updownarrow}_{#1}(#2)}
\newcommand{\ParikhMax}[2]{\alpha^\uparrow_{#1}(#2)}
\newcommand{\ParikhMin}[2]{\alpha^\downarrow_{#1}(#2)}
\newcommand{\ParikhVar}[2]{{\alpha}_{#1}(#2)}

\newcommand{\EqSet}[0]{\mathcal{E}}
\newcommand{\IntSet}[0]{\mathcal{I}}
\newcommand{\EqCnstr}[1]{\EqSet(#1)}
\newcommand{\IntCnstr}[1]{\IntSet(#1)}

\newcommand{\SMTB}{\textsc{IntSolver}}

\newcommand{\StrAt}[2]{\operatorname{strAt}(u, i)}

\allowdisplaybreaks

\begin{document}

\title{On Solving String Equations via Powers and Parikh Images}
\author{Clemens Eisenhofer\inst{1}\textsuperscript{(\Envelope)}\orcidID{0000-0003-0339-1580}
\and Theodor Seiser\inst{1}\orcidID{0009-0004-6422-7025} 
\and Nikolaj Bj{\o}rner\inst{2}\orcidID{0000-0002-1695-2810} \and Laura~Kov\'acs\inst{1}\orcidID{0000-0002-8299-2714}
}
\authorrunning{Eisenhofer et al.}
\institute{TU Wien, Vienna, Austria \\
\email{clemens.eisenhofer@tuwien.ac.at}
\and Microsoft Research,  Redmond, USA}

\maketitle


\begin{abstract}
We present a new approach for solving string equations as extensions of Nielsen transformations. Key to our work are the combination of three techniques: a power operator for strings; generalisations of Parikh images; and equality decomposition.
Using these methods allows us to solve complex string equations, including less commonly encountered SMT inputs over strings. 

\end{abstract}

\section{Introduction}
String solving is used in a wide range of applications, including formal verification~\cite{Norn2,Ostrich}, security analysis~\cite{DBTestInputGeneration,BlackOstrich,S3,WebAnalysis}, and automated reasoning~\cite{StringSolvingSurvey,Mosca}.
Modern Satisfiability Modulo Theories (SMT) solvers~\cite{SMT}, such as \textsc{Z3}~\cite{Z3}, \textsc{cvc5}~\cite{cvc5}, or \textsc{Princess}~\cite{princess}, support string constraints via specialised solving techniques for handling length constraints~\cite{Nielsen}, regular expressions~\cite{RegularWithLengths,SymDerivatives}, containment predicates~\cite{ExtStringCnstr}, and many others. However, solving string (word) equations that involve long repeated subsequences or contain mutually dependent string variables remains very challenging for state-of-the-art solvers~\cite{Norn2,Ostrich,Noodler,woorpje,cvc4StringSolving,z3Str4,Z3,S3}.
\begin{example}[Motivating Example]
    \label{ex:running}
    Consider the conjunction of the following two string equations: 
    \begin{align*}
        &x_3x_3x_4bx_5b &&\simeq&& x_5x_5x_5x_5x_4bb, \tag{$E_1$}\label{running:E1} \\
        &x_1x_1acx_2x_4x_2x_5 x_3bax_5x_3x_4x_3 &&\simeq&& x_2x_2abcx_1x_1x_3x_3 x_3x_4x_4ax_4 \tag{$E_2$}\label{running:E2}
    \end{align*}
    where $\simeq$ denotes (first-order) equality, $x_1, x_2, x_3, x_4, x_5$ are string variables and $a,b,c$ are constant characters. Solving such equations requires reasoning about string variables depending on themselves or mutually on each other needs, which is mostly out of scope of current techniques.
\end{example}
Our work addresses challenges similar to solving equations \eqref{running:E1}--\eqref{running:E2}. Our approach uses \emph{rewriting and generating rules over string equations} (Table~\ref{tab:rules}) as Nielsen transformations (Section~\ref{sec:smt}) over Nielsen graphs~\cite{Mosca,Makanin,Nielsen}.
We extend Nielsen transformation rules with:
\begin{description}
    \item[Equality decomposition] -- to decompose string equations during reasoning into subequations. Using equality decomposition greatly increases the applicability of string power reasoning and Parikh images.  
    \item[Explicit power representation] -- to solve string equations in which string variables depend on themselves. Furthermore, it allows us to compactly reason on long, but repetitive, strings (Section~\ref{sec:pwIntr}).

    \item[Generalized Parikh images] -- to 
     detect unsatisfiability when both Nielsen rules and string power reasoning fail (Section~\ref{sec:parikh}). We adjust generalised forms of Parikh images~\cite{Parikh} to string solving.
\end{description}

%
%
We implemented our framework as a prototype called \textsc{ZIPT} using the user-propagation framework~\cite{UserPropagator} of the SMT solver \textsc{Z3}~\cite{Z3}.
Our experiments demonstrate the practical potential of our method (Section~\ref{sec:experiments}) on \texttt{SMT-LIB2} benchmarks~\cite{SMTLIB} containing equations only. 

\section{Preliminaries}
\paragraph{Strings.} 
For our purposes, we will use an extended definition of a string term.
\begin{definition}[Token \& String Terms]
We fix a set $\Tokens$ of tokens and denote by $\Term$ the set of string terms. A \emph{string term} $u \in \Term$ is a finite sequence of \emph{tokens} $t \in \Tokens$ and we distinguish between 
\begin{itemize}
    \item \emph{concrete character} tokens, denoted as $a, b, c, d$, whose set is denoted by $\Alph$;
    \item \emph{symbolic characters} tokens, written as
    $o$, whose set is $\VarC$. 
    \item \emph{(string) variable} tokens, denoted as $x, y, z$, whose set is $\VarS$;
    \item \emph{power} tokens of the form $u^m$, where $u \in \Term$ does not contain string variables and $m$ is an arbitrary integer term potentially containing integer variables. We call $u$ the \emph{base} and $m$ the \emph{exponent} of a power token. 
\end{itemize}
\end{definition}
Our token notation might use indices, and we write $@ \in \Alph \cup \VarC$ to denote (concrete or symbolic) character tokens and $t \in \Term$ for a token of any kind. 
The sets $\Alph$, $\VarC$, and $\VarS$ are  mutually disjoint and $\Alph$ is additionally finite and non-empty. 
The set $\Term$ of string terms is the set of all finite sequences of tokens. Using regular expression notation, we have $\Term := \Tokens^*$ and reserve $\varepsilon$ for the empty sequence of tokens. $\varepsilon$ represents the neutral element of concatenation, and we denote concatenation by juxtaposing tokens.
The power token $u^m$ expresses that the token $u$ is repeated $m$ times; that is, $u^0 = \varepsilon$ and $u^m = uu^{m - 1}$ if $m > 0$ where $m$ is some integer term. Given how power terms are introduced, negative values for the exponents can be excluded.
We denote by $\Len{u}$ the \emph{length} with $u \in \Term$; clearly, $\Len{u}$ represents a natural number. We write $u^R$ to denote the reverse string term of $u$. For example, $(ax(ab)^m)^R = (ba)^mx^Ra$. Finally, we note that a symbolic character $o$ is equivalent to a string variable $x$ with $\Len{x} = 1$. We nonetheless use symbolic characters to have simpler definitions later on.

\paragraph{String equations and substitutions.} A \emph{string equation} is $u_1 \simeq u_2$ over string terms $u_1, u_2 \in \Term$; here, we refer to $u_1$ as the left-hand-side (LHS) of the equation, whereas $u_2$ is the right-hand-side (RHS) of the equation. 
String equations that only contain string variables and concrete characters are called \emph{plain}; string equations that might also contain symbolic characters and powers are called \emph{extended}. 
We use $\STSubStr{u}$ to denote the set of \emph{consecutive tokens} within $u$. $\SLen{u}$ denotes the number of tokens in $u$, where power tokens in $u$ are considered atomic; we call this the \emph{symbolic length} of $u$: $\SLen{\varepsilon} := 0$ and $\SLen{tv} := 1 + \SLen{v}$.  For example, in $u = (abc)^mxb$ we have $\STSubStr{u} = \left\{~ \varepsilon, a, b, c, (abc)^m, x, ab, bc, (abc)^mx, xb, abc, (abc)^mxb ~\right\}$ and $\SLen{u} = 3$.

String terms $u \in \Term$ containing no string variables are called \emph{ground}. For ground terms $u$ we thus have $\STSubStr{u} \cap \VarS = \emptyset$. Nonetheless, ground terms might contain power and symbolic character tokens.

A \emph{substitution} $\sigma$ maps (i) string variables $x \in \VarS$ to string terms $u \in \Term$;  and (ii) symbolic characters $o \in \VarC$ to characters $@ \in \Sigma \cup \VarC$.
The set of variables $\left\{ x \in \VarS \mid \sigma(x) \neq x \right\}$ is finite. Similarly, for symbolic characters.
We write $(x / u) \in \sigma$ to denote $\sigma(x) = u$. 
A substitution $\sigma$ is \emph{eliminating} a string variable $x$ if $(x / u) \in \sigma$ and $u$ is ground. 
The \emph{application of a substitution $\sigma$} to an expression $C$ is denoted by $C[\sigma]$, where $C$ is a string term or a (set of) string/integer (in)equations.

Substitutions $\sigma$ are assumed to be acyclic and fully extended: For any non-identity $(x / u) \in \sigma$ and $y \in \STSubStr{u}$ we have $(y / y) \in \sigma$. We write $m_1 \simeq m_2$ and $m_1 \le m_2$ for integer (in)equations over integer terms $m_1, m_2$. 

\paragraph{Interpretations and models.} Given a string equation $u \simeq v$, an \emph{interpretation} $\Int$ of this equation is a substitution together with an assignment of integer variables to integer values.
For every $x \in \VarS \cap (\STSubStr{u} \cup \STSubStr{v})$ we have $(x / w) \in \Int$, where $w \in \Alph^*$; further, for every $o \in \VarC \cap (\STSubStr{u} \cup \STSubStr{v})$ we have $(o / a) \in \sigma$ with $a \in \Alph$. 
An interpretation $\Int$ is called a \emph{plain model} of a set $S$ of plain string equations if the application of $\Int$ to $S$ results in a simplified set of equations of the form $u \simeq u$ with $u \in \Alph^*$. If $S$ has a model $\Int$, then $S$ is \emph{satisfiable}; otherwise $S$ is \emph{unsatisfiable}. Similarly, an \emph{extended model} of a set of extended string equations and integer (in)equalities simplifies every extended string equation to $u \simeq u$ with $u \in \Term^*$ and satisfies all integer constraints. All string variables and symbolic characters in $u$ are implicitly assumed to be universally quantified. Any extended model of a set of plain string equations can be made into a plain model by substituting all symbolic characters with some character of $\Alph$ and string variables with some element of $\Alph^*$. Power tokens can be completely unwound by using the integer value of their evaluated power term.

\section{String Solving -- Workflow}
\label{sec:smt}
\begin{figure}[t]
    \centering
    \includegraphics[width=.9\textwidth]{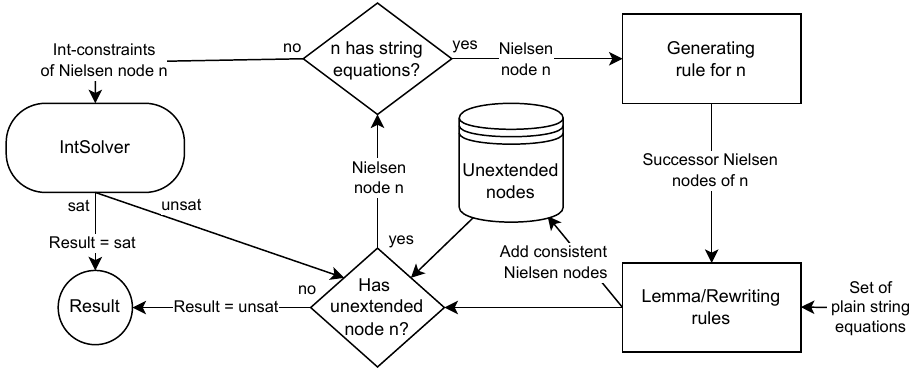}
    \caption{String solving workflow.}
    \label{fig:overview}
\end{figure}

Given a non-empty \emph{set $S$ of plain string equations, our task is to decide its satisfiability}.  
Doing so, we solve extended string equations corresponding to $S$, potentially containing symbolic characters and powers as well as integer (in)equalities.
To this end, we expand Nielsen graphs of string equations (Sections~\ref{sec:NielsenGraphs}) 
by repeatedly applying an extended set of Nielsen transformation rules (Sections~\ref{sec:extendenNR}). We simplify extended string equations and integer (in)equations, by extending Nielson transformations to arbitrary string terms $u \in \Term$ rather than only plain ones.
Our approach is summarised in Figure~\ref{fig:overview} and detailed in the following.
\subsection{Expansion of Nielsen Graphs}
\label{sec:NielsenGraphs}
We detect the satisfiability of a set of string equations $S$ with the help of Nielsen graphs~\cite{Nielsen}, which we expand and simplify using our Nielsen transformation rules from Section~\ref{sec:mainRules}. 

\paragraph{Nielsen graphs.} Simply put, we consider a \emph{(Nielsen) node} $n$ to be the tuple $\langle\EqCnstr{n}, \IntCnstr{n}\rangle$, where $\EqCnstr{n}$ is a finite set of extended string equations and $\IntCnstr{n}$ is its finite set of integer (in)equalities. In particular, the Nielsen node corresponding to the input set $S$ of plain string equations is $\langle S, \emptyset\rangle$; this node is the \emph{root node} of the Nielsen graph of $S$.

We assume that trivially satisfied constraints in $\EqCnstr{n}$ and $\IntCnstr{n}$, such as $\varepsilon \simeq \varepsilon$ or $0 \le 1$, are implicitly removed from $n$.  
Given some integer (in)equation $C$, we write $n \models C$ to denote that we can derive $C$ from $\IntCnstr{n}$. 

A Nielsen node $n$ has a finite number of successor nodes and is called \emph{inconsistent} if either (i) $\bot \in \EqCnstr{n}$, or (ii) $\IntCnstr{n} \models \bot$, or (iii) all successors nodes of $n$ are inconsistent; where $\bot$ denotes the always false formula. Contrary, $n$ is \emph{satisfied} if (i) $\EqCnstr{n} = \emptyset$, and (ii) $\IntCnstr{n} \not\models \bot$. For checking $n \models C$, we assume access to a sound integer reasoner $\SMTB$.

The node $n$ is \emph{extended} if its successor nodes are already added to the Nielsen graph, given that $n$ is neither satisfied nor inconsistent. Dually, a node that is not extended yet is called \emph{unextended}. A Nielsen graph captures dependencies between Nielsen nodes and, as such, witnesses the (un)satisfiability of constraints by a \emph{subgoal reduction}: if the constraints  $\EqCnstr{n} \cup \IntCnstr{n}$ of node $n$ have a model, then at least one of the successors of $n$ has a model as well.  
Nielsen graphs can thus be seen as a proof object or a variant of a tableau derivation. 
\paragraph{String Solving via Expansion of Nielsen Graphs.} We use Nielsen graphs to represent and manipulate our (initial) set $S$ of (plain) string equations. Doing so, we apply extended Nielsen transformation rules (Section~\ref{sec:extendenNR} and its extension in the succeeding sections) to simplify Nielsen nodes. In the sequel, let $\operatorname{simpl}(n)$ denote the simplified version of Nielsen node $n$.

Satisfiability of the input set $S$ of plain string equations is decided via recursively expanding the Nielsen graph of $S$, as shown in Figure~\ref{fig:overview}. We start by (i) adding the plain set of equations $S$ to a Nielsen graph in the form of an unextended and simplified Nielsen node $\operatorname{simpl}(\langle S, \emptyset\rangle)$ as the root of the Nielsen graph; here, we use the rewriting and lemma rules of Section~\ref{sec:mainRules}.
Next, (ii) we choose an arbitrary unextended Nielsen node $n$ of the Nielsen graph and generate its successors $n_1, \ldots, n_k$ via some generating rule from Section~\ref{sec:mainRules}. Then, (iii) we consider the simplified versions of the successor nodes $\operatorname{simpl}(n_i)$ of $n$ that are not inconsistent. These simplified successors of $n$ are added as new unexpanded nodes to the Nielsen graph by connecting $n_i$ with $n$ via a respective edge -- usually, we loop back to (ii). Once, (iv) a satisfied node $n$, reachable from the root node, is found, we report the initial set of equations $S$ to be satisfiable. Satisfiability of $S$ is implied by the construction of the expanded Nielsen graph: if a node $n$ is satisfied, all its string equations $\EqCnstr{n}$ have been removed and its remaining integer constraints $\IntCnstr{n}$ are satisfiable. On the other hand, (v) if there are no unextended Nielsen nodes reachable from the root node anymore and no satisfied node has been found, we conclude the unsatisfiability of the input $S$.

\begin{remark}[Termination of expanding Nielsen graphs] If a set of plain string equations $S$ is satisfiable, there exists at least one satisfied Nielsen node $n$ that is reachable from the root $\operatorname{simpl}(\langle S, \emptyset\rangle)$. Yet, a fully extended Nielsen graph might be infinitely large and might contain infinitely many satisfied nodes. Termination of unsatisfiable instances is not guaranteed.
As our rules (Section~\ref{sec:mainRules}) are usually not invertible -- $n$ being satisfiable implies only one successors to be satisfiable as well -- we require a fair tree traversal methods, such as breadth first or iterative deepening, in practice to make our approach terminate more robustly on satisfiable instances $S$.
\end{remark}

\subsection{Extended Nielsen Transformations Rules}\label{sec:extendenNR}
Our string solving expands Nielsen graphs of string equations, by rewriting and generating Nielsen nodes $n$ within the respective Nielsen graphs, as discussed in Section~\ref{sec:NielsenGraphs}. Namely, we simplify some node $n$ by $\operatorname{simpl}(n)$ via (i) lemma rules, (ii) term rewriting rules, and (iii) equation rewriting rules, in order to obtain the ``easier-to-handle" node $\operatorname{simpl}(n)$. We use (iv) generating rules to introduce the successor nodes $n'$ of $n$, where the constraints of $n'$ result from the application of a substitution $\sigma_{n'}$ to the constraints of $n$ and the addition of further constraints $C_{n'}$. That is, $n' = \operatorname{simpl}(\langle \EqCnstr{n}[\sigma_{n'}],~ \IntCnstr{n}[\sigma_{n'}] \cup C_{n'}\rangle)$. 

In the sequel, we fix an arbitrary Nielsen node $n$ and introduce the following rules relative to $n$. 
\subsubsection{Lemma rules.} 
If $\EqCnstr{n} \cup \IntCnstr{n}$ contains the power tokens $u^m$, string equations $u \simeq v$, or string variables $x$, we add the respective integer constraints $m \ge 0$, $\Len{u} \simeq \Len{v}$, and $\Len{x} \ge 0$ to $\IntCnstr{n}$.
\subsubsection{Term rewriting rules.}
\label{sec:rewriting} 
In the sequel, we write $e_1 \leadsto e_2$ to mean that $e_1$ is replaced by $e_2$ everywhere in the considered Nielsen node. 
Term rewriting rules are used to simplify length constraints and rewrite power terms, as follows: 
\begin{align*}
 &   \Len{uv} ~\leadsto~ \Len{u} + \Len{v} &~~~~& \Len{u^m} ~\leadsto~ m \Len{u}  &~~~~& 
   \varepsilon^m ~\leadsto~ \varepsilon \\
&  \Len{\varepsilon} ~\leadsto~ 0 && \Len{@} ~\leadsto~ 1 && 
v^m~ \leadsto~ \varepsilon, \text{~if } n \models m \simeq 0 \\
 &    \left(v^{m_1}\right)^{m_2} ~\leadsto~ v^{m_1m_2} && v^{m_1}v^{m_2} ~\leadsto~ v^{m_1+m_2} 
&&
    v^m ~\leadsto~ v,  \text{~if } n \models m \simeq 1
\end{align*}
A term is \emph{fully rewritten} if no more rewriting rule can be applied to it. An (in)equality constraint is \emph{fully rewritten} if all its terms are fully rewritten.
\begin{example}[Computing $\operatorname{simpl}(n)$ for Example~\ref{ex:running}]\label{ex:running:simpl}
    Consider the two input equations of Example~\ref{ex:running} and let $n$ be their respective node $n$ in the corresponding Nielsen graph. We drop the common $b$ suffix in~\eqref{running:E1} and use lemma rules to add a length constraints for each string equation which can be rewritten into the fully rewritten constraints $2\Len{x_{3}} \simeq 3\Len{x_{5}}$ and $2\Len{x_5} \simeq \Len{x_4}$, using respective term rewriting rules. Finally, we add the side constraints $\Len{x_i} \ge 0$ for all $1 \le i \le 5$. The resulting set of constraints defines $\operatorname{simpl}(n)$. 
\end{example}
\subsubsection{Equation rewriting and generating rules.}
\label{sec:mainRules} 

Equation rewriting rules remove some prefix of a string equation of $\EqCnstr{n}$, sometimes under additional side conditions.
In case no rewriting rule is applicable over $n$ anymore, a generating rule over $n$ is triggered to add additional information, which usually happens by considering multiple cases in the form of successor nodes.

Equation rewriting/generating rules for an extended string equation $u \simeq v \in \EqCnstr{n}$ are given based on the first tokens of $u$ and $v$. 
Our rewriting and generating rules are summarised in Table~\ref{tab:rules}, listed per possible combinations of first tokens, or $\varepsilon$, of the LHS and RHS (columns~1-2 of Table~\ref{tab:rules}). Primed variables $x'$ in the table denote fresh string variables. A rule can be applied if the equation is of the given form and the side condition is satisfied in the current Nielsen node.
We note that we can also apply all rules to $v \simeq u$, $u^R \simeq v^R$, and $v^R \simeq u^R$.
Finally, any equation rewritten to $\varepsilon \simeq \varepsilon$ is removed from $\EqCnstr{n}$, and similarly any trivially satisfied integer (in)equation.

Equation rewriting rules replace some equation $u \simeq v$ either by a new equation or detect a conflict $\bot$ (column~4), potentially requiring some side condition to be satisfied (column~3).
On the other hand, generating rules apply substitutions (column~5) to $n$ in order to generate successor node(s) $n'$ of $n$ or add additional new (integer) constraints to $n'$ (column~6) to make further equation rewriting rules applicable.
{\small
\begin{table}[t]
    \centering
    {\small
    \begin{tabular}{|c|c||c|c||c|c|}
        \hline 
        \multirow{2}{*}{LHS} & \multirow{2}{*}{RHS} & \multicolumn{2}{|c|}{Rewriting rules} & \multicolumn{2}{|c|}{Generating rules} \\
        \cline{3-6}
        & & Side Cond. & New Equation & Substitution & New Cnstr.\\
        \hline
        $tu$ & $tv$ & -- & $u \simeq v$ & -- & -- \\
        \hline
        $\varepsilon$ & $av$ & -- & $\bot$ & -- & -- \\
        \hline
        $\varepsilon$ & $ov$ & -- & $\bot$ & -- & -- \\
        \hline
        $\varepsilon$ & $xv$ & -- & -- & $x / \varepsilon$ & -- \\
        \hline
        \multirow{2}{*}{$\varepsilon$} & \multirow{2}{*}{$w^mv$} & \multirow{2}{*}{--} & \multirow{2}{*}{--} & -- & $\{ w \simeq \varepsilon \}$ \\
        & & & & -- & $\{ m \simeq 0 \}$ \\
        \hline
        $au$ & $bv$ & -- & $\bot$ & -- & -- \\
        \hline
        $ou$ & $@v$ & -- & -- & $o / @$ & -- \\
        \hline
        \multirow{2}{*}{${}^{\dagger} @u$} & \multirow{2}{*}{$xv$} & \multirow{2}{*}{--} & \multirow{2}{*}{--} & $x / \varepsilon$ & -- \\
        & & & & $x / @x'$ & -- \\
        \hline
        \multirow{2}{*}{${}^\ddagger @u$} & \multirow{2}{*}{$w^mv$} & $n \models m \simeq 0$ & $@u \simeq v$ & -- & $\{ m \simeq 0 \}$ \\
        & & $n \models m > 0$ & $@u \simeq ww^{m - 1}v$ & -- & $\{ m > 0 \}$ \\
        \hline
        \multirow{4}{*}{$xu$} & \multirow{4}{*}{$yv$} & \multirow{4}{*}{--} & \multirow{4}{*}{--} & $x / \varepsilon$ & -- \\
        & & & & $y / \varepsilon$ & $\{ \Len{x} > 0 \}$ \\
        & & & & $y / xy'$ & $\{ \Len{x} > 0 \}$ \\
        & & & & $x / yx'$ & $\{ \Len{y} > 0, \Len{x'} > 0 \}$ \\
        \hline
        \multirow{4}{*}{${}^{\dagger} xu$} & \multirow{4}{*}{$w^mv$} & \multirow{4}{*}{--} & \multirow{4}{*}{--} & $x / w^mx'$ & -- \\
        & & & & $x / w^{m'} p_1$ & $s_1 \cup \{ 0 \le m' < m \}$ \\
        & & & & \ldots & \ldots \\
        & & & & $x / w^{m'} p_k$ & $s_k \cup \{ 0 \le m' < m \}$ \\
        \hline
        \multirow{2}{*}{${}^\ddagger w^{m_1}u$} & \multirow{2}{*}{$w^{m_2}v$} & $n \models m_1 \ge m_2$ & $w^{m_1 - m_2}u \simeq v$ & -- & $\{ m_1 \ge m_2 \}$ \\
        & & $n \models m_1 < m_2$ & $u \simeq w^{m_2 - m_1}v$ & -- & $\{ m_1 < m_2 \}$ \\
        \hline
        \multirow{2}{*}{${}^\ddagger w_1^{m_1}u$} & \multirow{2}{*}{$w_2^{m_2}v$} & $n \models m_2 \simeq 0$ & $w_1^{m_1}u \simeq v$ & -- & $\{ m_2 \simeq 0 \}$ \\
        & & $n \models m_2 > 0$ & $w_1^{m_1}u \simeq w_2w_2^{m_2 - 1}v$ & -- & $\{ m_2 > 0 \}$ \\
        \hline
    \end{tabular}}\smallskip
    \caption{Equation rewriting rules and generating rules for an extended equation $u \simeq v$.}
    \label{tab:rules}
\end{table}
}
\begin{example}[Generating rules for $\operatorname{simpl}(n)$ in Example~\ref{ex:running}]
Consider the simplified node 
$\operatorname{simpl}(n)$ computed in Example~\ref{ex:running:simpl} for Example~\ref{ex:running}. 
The set of applicable generating rules is given by the LHS/RHS token combinations of  $x_3$ and $x_5$;  $x_5^R$ and $b$;  $x_1$ and $x_2$; and  $x_3^R$ and $x_4^R$. 
\end{example}


We stress that expanded Nielsen graphs using the rules of Table~\ref{tab:rules} correspond to a proof ``tree'', as illustrated next. 

\begin{example}[A fully expanded Nielsen graph for $xx \simeq yb$]
    Consider the plain string equation $xx \simeq yb$. We establish its satisfiability by constructing its expanded Nielsen graph, as presented in Section~\ref{sec:NielsenGraphs} and using the rules of Table~\ref{tab:rules}. The fully expanded Nielsen graph of $xx \simeq yb$ is shown in Figure~\ref{fig:nielsenGraph}. By reusing variable names, a single node could be used rather than multiple isomorphic ones. Hence, $x' \simeq \varepsilon \wedge \Len{x'} \simeq 0$ and $x'' \simeq \varepsilon \wedge \Len{x''} \simeq 0$ could be contracted to the same node.
    \begin{figure}[t]
        \centering
        \includegraphics[width=\textwidth]{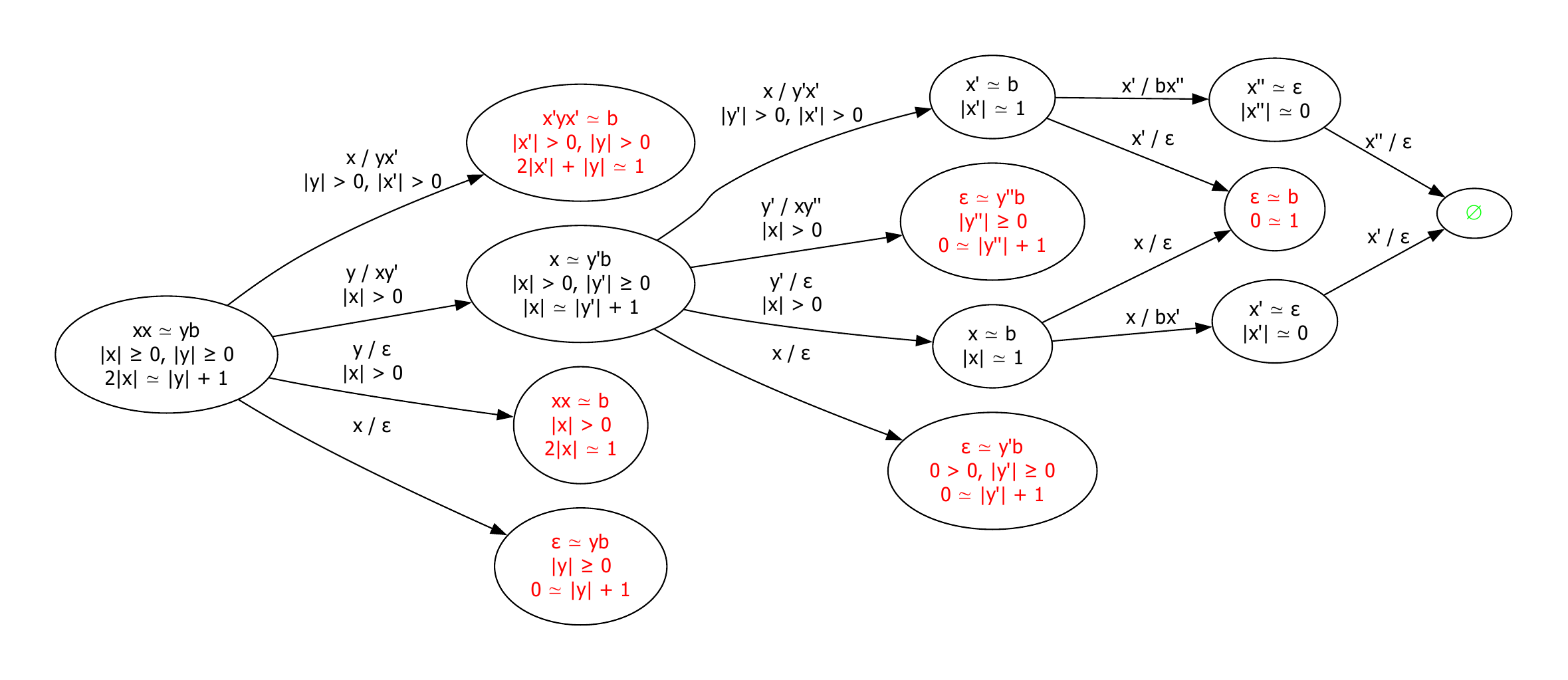}
        \caption{Fully expanded Nielsen graph for the plain string equation $xx \simeq yb$, representing a variant of a tableau derivation.}
        \label{fig:nielsenGraph}
    \end{figure}
\end{example}

We conclude this section by noting that the rules of Table~\ref{tab:rules} annotated by ${}^\dagger$ and ${}^\ddagger$ require special care, as discussed in Section~\ref{sec:pwIntr}. In particular, the case with $xu$ on the LHS and $w^mv$ on the RHS requires $k + 1$ branches given by the finite set $\SDec{w} = \{ \langle p_1, s_1\rangle, \ldots, \langle p_k, s_k\rangle \}$ which denotes the set of all syntactic prefixes $p_i$ of $w$ together with potential side conditions $s_i$. The set $\SDec{w}$ is detailed in Section~\ref{sec:pwIntr}.
We optimise the application of those rules by using more extended rules (Sections~\ref{sec:eqSplit}--\ref{sec:parikh}) discussed next. 

\section{Equality Decomposition}
\label{sec:eqSplit}
This section provides a remedy for some cases when string equations cannot be solved using the Nielsen transformation rules of Section~\ref{sec:mainRules}. 
Intuitively, we decompose string equations into subequations that can be further solved using our transformation rules. 
Note that the rules of Table~\ref{tab:rules} can only be applied upon the first (last) tokens of the LHS/RHS of string equations. With equality decomposition, we split equations into smaller ones. As such, the equality decomposition enables us to apply Table~\ref{tab:rules} on tokens at positions different from the first tokens of LHS/RHS. 

In a nutshell, we proceed as follows. Consider an extended string equation $u_1u_2 \simeq v_1v_2 \in \EqCnstr{n}$ and let $d = \Len{u_1} - \Len{v_1}$ be a known integer constant.
Equality decomposition is applied by splitting the equation into two equations, potentially padding one side each by $d$ characters. 
If $d = 0$, the equation $u_1u_2 \simeq v_1v_2$ can be decomposed into the set $\left\{ u_1 \simeq v_1, u_2 \simeq v_2 \right\}$ of two smaller equations. 
If all string terms $u_1, u_2, v_1, v_2$ contain at least one string variable or a power token, we replace $u_1u_2 \simeq v_1v_2$ in $\EqCnstr{n}$ by $\left\{ u_1 \simeq v_1, u_2 \simeq v_2 \right\}$. 
For example, we decompose $xayw \simeq ybxz$ into $xay \simeq ybx$ and $w \simeq z$, as $\Len{xay} - \Len{ybx} = 0$.

On the other hand, if $d > 0$, then decomposing $u_1u_2 \simeq v_1v_2$ comes with the additional requirement that the last $d$ characters of $u_1$ are the same as the first $d$ characters of $v_2$. To do so, we introduce $\Len{d}$ fresh symbolic character tokens $\bar{o_d}' = o_1', \ldots, o_d'$ 
to represent these $d$ characters of $u_1$. Note again, this is equivalent to introducing one fresh string variable $x$ with $\Len{x} = d$. Then, $u_1u_2 \simeq v_1v_2$ is decomposed into the equations: 
\[ \left\{~ u_1 \simeq v_1 \bar{o}_d',~ \bar{o}_d' u_2 \simeq v_2 ~\right\}. \]
Our equality decomposition approach thus uses length constraints $\Len{u_1} - \Len{v_1}$, which can be facilitated by the integer constraints $\IntCnstr{n}$ of $n$. Our approach can be seen as a strengthened variant of the usual decomposition rule without padding, primarily used as a preprocessing rule in other solvers:
We use $d$ symbolic character tokens for padding string equations in order to enable their decomposition.
\begin{example}[Equality decomposition in Example~\ref{ex:running}]
    \label{ex:runningAfterEqSplit}
    Using the lemma and term rewriting rules of Example~\ref{ex:running:simpl}, we derive: 
    \[ 2\Len{x_{3}} \simeq 3\Len{x_{5}} \wedge 2\Len{x_5} \simeq \Len{x_4}. \]
    Consequently, we infer $d=\Len{x_1x_1acx_2x_4x_2x_5} - \Len{x_2x_2abcx_1x_1x_3x_3} = 1$. 
    We decompose \eqref{running:E2} using a single symbolic character $o$ and replace ~\eqref{running:E2} by the resulting two new equations \eqref{running:E3} and \eqref{running:E4}: 
    \begin{align*}
        &x_1x_1acx_2x_4x_2x_5o &&\simeq&& x_2x_2abcx_1x_1x_3x_3, \tag{$E_3$}\label{running:E3} \\
        &x_3bax_5x_3x_4x_3 &&\simeq&& ox_3x_4x_4ax_4 \tag{$E_4$}\label{running:E4}
    \end{align*}
    In other words, \eqref{running:E2} is replaced by \eqref{running:E3} and \eqref{running:E4} in $\EqCnstr{n}$, and we further simplify $n$ by lemma and term rewriting rules.
\end{example}

\section{Ground Power Introduction}
\label{sec:pwIntr}
We introduce power terms to eliminate lengthy and repetitive substrings in string equations, and compress sequences of rewriting and generating rule applications into power terms. 
Power terms allow us to eliminate a potentially unbounded number of equation decompositions when applying substitutions of the form $x / @x'$; here, $x'$ is again a fresh string variable. For example, when $xu = axv$ is satisfiable, all its models need $x$ to be $a^m$, for some $m \ge 0$. Finding such models using equality decomposition in combination with the rules of Table~\ref{tab:rules} requires analysing infinitely many token combinations for some unsatisfiable equations. Even in the case of satisfiable problem instances, the required analysis can become very long. In order to circumvent this, we use rules based on the property that $xu = vx$, with $v \ne \varepsilon$, implies $x = (w_1w_2)^mw_1$, $v = w_1w_2$, and $u = w_2w_1$ for some natural number $m$ and words $w_1$ and $w_2$~\cite{lyndon1962equation}.

\paragraph{Power introduction and rules ${}^\dagger$  of Table~\ref{tab:rules}.} The rules of Table~\ref{tab:rules} that are annotated by ${}^\dagger$ are only applied when power introduction is not applicable. In such cases, given a ground string term $w \neq \varepsilon$ and constraint $xu \simeq wxv$, we have: 

\begin{equation}\label{eq:powerIntroDager}
    \exists m. \bigvee_{\langle p, s \rangle \in \SDec{w}} \left(x \simeq w^m p \wedge \bigwedge s\right) \vee w \simeq \varepsilon
\end{equation}
where $\SDec{w}$ denotes the set of pairs $\langle p, s \rangle$ of possible strict prefixes $p$ of $w$, such that there is only a finite set $s$ of side conditions to constrain $p$. Formally, $\SDec{w}$ is defined for a ground string term $w$ as: 

\begin{itemize}
    \item $\SDec{uv} := \SDec{u} \cup \{ \langle up, s \rangle \mid \langle p, s \rangle \in \SDec{v} \}$, 
    \item $\SDec{\varepsilon} := \emptyset$,
    \item $\SDec{@} := \{ \langle \varepsilon, \emptyset\rangle \}$, 
    \item $\SDec{u^m} := \{ \langle u^{m'} p, \{ 0 \le m' < m \} \cup s \rangle \mid \langle p, s \rangle \in \SDec{u} \}$
\end{itemize}
where $m'$ is a fresh integer variable. 

Let us make the following observation on the interplay between power introductions and prefix $\SDec{w}$  analysis. When considering equations $xu \simeq wxv$, an unsoundness problem may arise when $w$ represents $\varepsilon$ in a potential model, as $w$ would have cancelled out. Such cases may happen when $w$ consists of power tokens all becoming $\varepsilon$ because their exponents are zero. Based on the definition of $\SDec{w}$, we would introduce a fresh $m'$ integer variable with $0 \le m' < m$, which yields a conflict when $m = 0$ and thus falsely eliminates a model. As a remedy, we also explicitly split upon $w \simeq \varepsilon$ to cover this case. On the other hand, if $w$ is not $\varepsilon$, power introduction allows us to eliminate string variables $x$ by replacing them with the help of integer constraints, as shown next.

\begin{example}[$xbxa \simeq axbx$]
    Without introducing powers, applying the rules of Table~\ref{tab:rules} would fail to solve the unsatisfiable equation $xbxa \simeq axbx$. Applying the rules from Table~\ref{tab:rules} leads to an infinite chain of substitutions of the form $x / ax'$, resulting in increasingly longer string terms.
    By introducing power terms, we eliminate $x$ via the single substitution $x / a^m$: applying this substitution to $xbxa \simeq axbx$ yields $a^mba^ma \simeq aa^mba^m$ that can be easily simplified to $\bot$.
\end{example}

\paragraph{Power introduction and rules ${}^\ddagger$ of Table~\ref{tab:rules}.} We next comment on the use of power introduction together with the Nielsen transformation rules 
annotated by ${}^\ddagger$ in Table~\ref{tab:rules}. 
When the first token of some side of an equation is a power token, we first check whether the other side of the equation can be rewritten so that it starts with a power token with the same base, so that they can cancel out. For doing so, we apply additional term rewriting rules: 
\begin{align*}
    ww^m &&\leadsto&& w^{m+1} &\qquad~& w_2\left(w_1w_2\right)^m &&\leadsto&& \left(w_2w_1\right)^mw_2\\
    w^mw &&\leadsto&& w^{m+1} &\qquad~& \left(w_1w_2\right)^m w_1 &&\leadsto&& w_1 \left(w_2w_1\right)^m\\
    ww &&\leadsto&& w^2 
\end{align*}
These additional rules considerably reduce the number of equational decompositions when introducing powers. However, more importantly, they are necessary to handle equations like $a(ba)^mu \simeq (ab)^mv$. In case the solver does not realize that the prefixes can be rewritten into the same power, the rules of Table~\ref{tab:rules} would suggest an explicit unwinding for each possible value of $m$.

\paragraph{Strengthening power reductions.}
We generalize the rule of~\eqref{eq:powerIntroDager} over $xu \simeq wxv$ to be applicable over sets of equations $EQ \subseteq \EqCnstr{n}$ given by: 

\[ EQ := \left\{~ w_1x_k u_1 \simeq x_1 v_1,~ w_2x_1 u_2 \simeq x_2 v_2,~ \ldots,~ w_kx_{k-1} u_k \simeq x_k v_k ~\right\},\]
with $k\geq 1$, and all terms $w_1, \ldots, w_k$ being ground. 
Similarly to \eqref{eq:powerIntroDager}, the 
set $EQ$ enforces the more general form of power introduction
\begin{equation}
    \label{eq:powerIntroFull}
    \exists m. \bigvee_{1 \le i \le k\vphantom{\langle\rangle}}~\bigvee_{\langle p, s \rangle \in \SDec{\bar{w}_i}} \left(x_i \simeq \bar{w}_i^m p \wedge \bigwedge s\right) \vee \bar{w}_1 \simeq \varepsilon,
\end{equation}
where $\bar{w}_i := w_i w_{i-1} \ldots w_1 w_k \ldots w_{i + 1}$.

\begin{example}[Power introduction in Example~\ref{ex:runningAfterEqSplit}]
    \label{ex:runningAfterPowers}
    After decomposing \eqref{running:E2} into \eqref{running:E3} and \eqref{running:E4}, we introduce a power token that is justified by the prefix of \eqref{running:E4}: the LHS of \eqref{running:E4} starts with $x_3$ and the RHS with $ox_3$. As the string term consisting only of the symbolic character $o$  is ground and cannot be $\varepsilon$, we apply power introduction and consider only one successor node as $\SDec{o} = \{ \langle \varepsilon, \emptyset \rangle \}$: the successor is generated by $x_3 / o^{m_1}$ for some fresh constant $m_1$. From a lemma rule, we get $m_1 \ge 0$ and we rewrite the length constraints $\Len{x_3[x_3 / o^{m_1}]}$ to $m_1$. The string equations \eqref{running:E1}, 
    \eqref{running:E3} and \eqref{running:E4}  respectively become: 
    \begin{align*}
        &o^{2m_1}x_4bx_5 &&\simeq&& x_5x_5x_5x_5x_4b, \tag{$E_1'$}\label{running:e1'} \\
        &x_1x_1acx_2x_4x_2x_5o &&\simeq&& x_2x_2abcx_1x_1o^{2m_1}, \tag{$E_3'$}\label{running:e3'}  \\
        &o^{m_1}bax_5o^{m_1}x_4o^{m_1} &&\simeq&& oo^{m_1}x_4x_4ax_4 \tag{$E_4'$}\label{running:e4'}.
    \end{align*}
    Thanks to the additional rewrite rules introduced previously in this section, token $oo^{m_1}$ on the RHS of \eqref{running:e4'} can be rewritten to $o^{m_1 + 1}$. As such, common prefixes of \eqref{running:e3'} are removed, and we apply $o / b$. We get the respective equations of \eqref{running:e1'}, \eqref{running:e3'} and \eqref{running:e4'} as:
    \begin{align*}
        &b^{2m_1}x_4bx_5 &&\simeq&& x_5x_5x_5x_5x_4b, \tag{$E_1''$} \label{running:e1''} \\
        &x_1x_1acx_2x_4x_2x_5b &&\simeq&& x_2x_2abcx_1x_1b^{2m_1}, \tag{$E_3''$} \label{running:e3''} \\
        &ax_5b^{m_1}x_4b^{m_1} &&\simeq&& x_4x_4ax_4. \tag{$E_4''$}\label{running:e4''} 
    \end{align*}
    Further powers can be introduced, either by using \eqref{running:e4''} because of the suffix $x_4 b^{m_1}$ on the LHS and $x_4$ on the RHS, or via the generalised form of power introduction \eqref{eq:powerIntroFull} using equations \eqref{running:e1''} and \eqref{running:e4''}.
    In the latter case, we end up with the four successor nodes given by:
    \[ \left\{~ x_4 / (ab^{2m_1})^{m_2},~ x_4 / (ab^{2m_1})^{m_2} ab^{m_3},~ x_5 / (b^{2m_1}a)^{m_2} b^{m_3}, ~ x_5 / (b^{2m_1}a)^{m_2} b^{2m_1} ~\right\} \] 
    and $m_3 < 2m_1$. 
    In the former case, we have two cases: applying $x_4 / b^{m_3} (b^{m_1})^{m_2}$ with $m_3 < m_1$ -- which simplifies to $b^{m_1m_2 + m_3}$ -- or adding $b^{m_1} \simeq \varepsilon$.
    We chose the former case and consider the first successor first: applying $x_4 / b^{m_3} (b^{m_1})^{m_2}$.
    \begin{align*}
        &b^{2m_1 + m_1m_2 + m_3}bx_5 &&\simeq&& x_5x_5x_5x_5b^{m_1m_2 + m_3}b, \tag{$E_1'''$} \label{running:e1'''} \\
        &x_1x_1acx_2b^{m_1m_2 + m_3}x_2x_5b &&\simeq&& x_2x_2abcx_1x_1b^{2m_1}, \tag{$E_3'''$} \label{running:e3'''} \\
        &ax_5b^{2m_1} &&\simeq&& b^{2m_1m_2 + 2m_3}a. \tag{$E_4'''$} \label{running:e4'''}
    \end{align*}
    As $0 \le m_3 < m_1$, we derive $m_1 > 0$ and thus we unwind $b^{2m_1}$ to $b^{2m_1 - 1}b$ in \eqref{running:e4'''} using the respective rule from Table~\ref{tab:rules}, resulting in a conflict. We therefore consider the successor node generated by adding the constraint $b^{m_1} \simeq \varepsilon$. This implies $m_1 = 0$, eliminating all power tokens and allowing us to derive $\Len{x_4} = \Len{x_5} = 0$. After further simplifications, we derive $x_4 / \varepsilon$ and $x_5 / \varepsilon$. As such, we end up with the Nielsen node containing the single plain string equation:
    \[ x_1x_1acx_2x_2b \simeq x_2x_2abcx_1x_1. \]
\end{example}

\section{Parikh Image}
\label{sec:parikh}
Parikh images~\cite{Parikh} can be an effective method to prove sets of string constraints to be unsatisfiable by extracting constraints on the number of occurrences of terminal characters~\cite{ParikhRegex}. It represents an abstraction of string equations that elides information about character positions, retaining only the number of occurrences. In the following, we establish a method that retains information about the relative positions of characters, resulting in a tighter abstraction. 
Our method computes a value we call \emph{(error-bounded) multi-sequence Parikh image}.

To formalise the Parikh images, let $\pi_a(u)$ be a function that represents the set of positions on which there is an $a \in \Alph$ in $u$.
The set $\{~ (a, \pi_a(u)) \mid a \in \Alph ~\}$ is a complete representation of $u$.
Let $\Parikh{a}{u} := \Len{\pi_a(u)}$, then an equation $u \simeq v$ can be over-approximated by $\bigwedge_{a \in \Alph}\Parikh{a}{u} \simeq \Parikh{a}{v}$. 
For example, $\Parikh{a}{xay} = \Parikh{a}{x} + 1 + \Parikh{a}{y}$ and $\Parikh{a}{xby} = \Parikh{a}{x} + \Parikh{a}{y}$, so the equation $xay \simeq ybx$ can be recognised as unsatisfiable immediately. We note that neither equality decomposition nor ground power introduction combined with the rules of Table~\ref{tab:rules} can detect this equation as unsatisfiable. 
Still, Parikh constraints based on single concrete characters fail to capture the infeasibility of equations such as the equation in the only Nielsen node of our running example: $x_1x_1acx_2x_2b \simeq x_2x_2abcx_1x_1$. Our method is going to accomplish this.

\subsection{Parikh Images for Unbordered Patterns}
In the following, we consider Parikh images for strings $w \in \Alph^+$ of length greater than $1$ -- we call these $w$ \emph{patterns} -- and assume that we have plain string equations. For the scope of this paper, we consider symbolic characters and power tokens as string variables when computing the Parikh image.
For example, in the plain equation $xabcy \simeq ybacx$, pattern $bc$ occurs on the LHS of the equation $\Parikh{bc}{xabcy} = \Parikh{bc}{x} + 1 + \Parikh{bc}{y}$ times and $\Parikh{bc}{ybacx} = \Parikh{bc}{y} + \Parikh{bc}{x}$ times on the RHS.
The equation is therefore unsatisfiable. Our running example also offers an opportunity to use such a pattern:
\begin{example}[Parikh image in Example~\ref{ex:runningAfterPowers}]
    \label{ex:runningAfterParikh}
    Consider the equation in our running Example~\ref{ex:runningAfterPowers}. We can consider pattern $w = abc$ and thus get $\Parikh{w}{x_1x_1acx_2x_2b} \simeq \Parikh{w}{x_2x_2abcx_1x_1}$. 
    This equation can be intuitively rewritten to $\Parikh{w}{x_1x_1} + \Parikh{w}{x_2x_2} \simeq \Parikh{w}{x_2x_2} + 1 + \Parikh{w}{x_1x_1}$ and further to $0 = 1$ which proves the equation itself, as well as the overall running example unsatisfiable, as all successor Nielsen nodes of the root node are inconsistent.
    In the following, we will formally define this ``intuitive rewriting''.
\end{example}

In contrast to usual Parikh images over singleton characters, there is no general notion of $\Parikh{w}{u}$ that decomposes over string constants and string variables in $u$.
For example, we cannot decompose $\Parikh{ab}{ybacx}$ into a sum over the components of $ybacx$ because the value depends on whether $x$ ends with $a$ or not.

We address this limitation by 
\begin{enumerate}
    \item Restricting Parikh images to character sequence patterns $w$ such that no proper suffix of $w$ is a prefix of $w$. Such words are called \emph{unbordered}.
    \item Defining over- $\ParikhMax{w}{u}$ and under-approximations $\ParikhMin{w}{u}$ such that unsatisfiability of a string equation $u \simeq v$ can be established when $\ParikhMax{w}{u} - \ParikhMin{w}{v}$ or, symmetrically, $\ParikhMax{w}{v} - \ParikhMin{w}{u}$ rewrite to a negative constant.
\end{enumerate}

Our rewriting rules for $\ParikhMax{w}{u}$ and $\ParikhMin{w}{u}$ satisfy the following properties:
\begin{lemma}[Correctness of Parikh images for Unbordered Patterns]
Assume $w$ is an unbordered pattern.
\begin{itemize}
    \item If $u \in \Alph^*$ then $\ParikhMax{w}{u} = \ParikhMin{w}{u} = $ number of occurrences of $w$ in $u$.

    \item Assume $\ParikhMax{w}{u}$ rewrites to $k + \sum_x c_x \ParikhVar{w}{x}$ for natural numbers $c_x$ and $k$. Then for every substitution $\sigma$ from string variables to $\Alph^*$, $\ParikhMax{w}{u[\sigma]} \leq k + \sum_x c_x\ParikhMax{w}{x[\sigma]}$.
    A symmetric property holds for $\ParikhMin{w}{u}$.
\end{itemize}
\end{lemma}
It remains to define the over- and under-approximations.
As the examples suggested, the restriction to unbordered patterns means that there are
many cases where the under- and over-approximations coincide. 
\begin{definition}[$d$-gaps and $w \bowtie u$]
\begin{description}
\item[$d$-gaps]
We call a string term $u$ a \emph{$d$-gap} (with $d > 1$) iff it is of one of the following forms: $u = xvy$ (with $v \in \Alph^*$), $u = xv$, or $u = vy$ ($v \in \Alph^+$ in the latter two cases) and $0 < \SLen{u} \le d$.
\item[$w \bowtie u$]
If $w$ is a pattern and $u$ is a $\Len{w}$-gap, we write $w \bowtie u$ to denote that there can be crossing occurrences of $w$ within a $\Len{w}$-gap $u$. Formally, $w \bowtie u$ is true if $u = xw_2y$ and we can decompose $w = w_1w_2w_3$ ($\{ w_1, w_3 \} \ne \varepsilon$), otherwise it is false. Similarly, the cases $u = w_1y$ requires $w = w_1w_2$ ($w_2 \ne \varepsilon$) and in cases $u = xw_2$ requires $w = w_1w_2$ ($w_1 \ne \varepsilon$)
\end{description}
\end{definition}
Using the definition of a gap, apply the following term rewriting rules to rewrite stepwise $\ParikhMin{w}{u}/\ParikhMax{w}{u}$ into the desired grouping for an unbordered pattern $w$: 
\begin{align*}
    \ParikhBoth{w}{x} & \leadsto \ParikhVar{w}{x} \\
    \ParikhBoth{w}{w_2} &\leadsto 0 && \text{if $w_2 \in \Alph^* \wedge \Len{w_2} < \Len{w}$} \\    
    \ParikhBoth{w}{uwv} &\leadsto 1 + \ParikhBoth{w}{u} + \ParikhBoth{w}{v} && \\
    \ParikhBoth{w}{ua_1w_2a_2v} &\leadsto \ParikhBoth{w}{ua_1w_2} + \ParikhBoth{w}{w_2a_2v} && \text{if $a_1w_2a_2 \in \Alph^+ \wedge \vphantom{A}$}\\
    & && ~~~~~\Len{a_1w_2a_2} = \Len{w} \wedge a_1w_2a_2 \ne w \\
    \ParikhBoth{w}{uxw_2yv} &\leadsto \ParikhBoth{w}{uxw_2} + \ParikhBoth{w}{w_2yv} && \text{if $xw_2y$ is $\Len{w}$-gap and $w \not\bowtie xw_2y$} \\
    \ParikhBoth{w}{aw_2yv} &\leadsto \ParikhBoth{w}{w_2yv} && \text{if $aw_2y$ is $\Len{w}$-gap and $w \not\bowtie aw_2y$} \\
    \ParikhBoth{w}{uxw_2a} &\leadsto \ParikhBoth{w}{uxw_2} && \text{if $xw_2a$ is $\Len{w}$-gap and $w \not\bowtie xw_2a$}
\end{align*}
We now define under- and over-approximations for cases not covered by the previous exact rewrite rules. They are:
\begin{align*}
    \ParikhMin{w}{x} &\leadsto \ParikhVar{w}{x} %
    & \ParikhMax{w}{x} &\leadsto \ParikhVar{w}{x} && \\
    \ParikhMin{w}{w_2u} &\leadsto \ParikhMin{w}{u} %
    & \ParikhMax{w}{w_2xu} &\leadsto 1 + \ParikhMax{w}{xu} && \text{if $w_2 \in \Alph^+$} \\
    \ParikhMin{w}{xw_2} &\leadsto \ParikhVar{w}{x} %
    & \ParikhMax{w}{xw_2} &\leadsto 1 + \ParikhVar{w}{x} && \text{if $w_2 \in \Alph^+$}\\
    && \ParikhMax{w}{xw_2yv} &\leadsto 1 + \ParikhVar{w}{x} + \ParikhMax{w}{yv} && \text{if $w_2 \in \Alph^*$}
\end{align*}
The correctness of the rewrite rules relies on first applying the non-approximate rules exhaustively before considering the cases presented by the approximations. Informally speaking, we require all $u$ in the remaining $\Parikh{w}{u}$ to be ``concatenations of $\Len{w}$-gaps'' that can contain crossing occurrences. 

Our procedure for filtering unsatisfiable equations $u \simeq v$ is now as follows:

\noindent{\bf (i)} Enumerate maximal unbordered patterns $w \in \Alph^*$ occurring in $u$ and $v$. The patterns are maximal only w.r.t. the pattern within the considered side $u$ or $v$.\smallskip

\noindent {\bf (ii)} For each of these $w$ rewrite $\ParikhMax{w}{u} - \ParikhMin{w}{v}$ to a sum $k + \sum_x c_x \ParikhVar{w}{x}$. If each $c_x$ is $0$ and $k < 0$, we can conclude the equation $u \simeq v$ to be unsatisfiable. 

\begin{example}[$xaxaabbby \simeq xyabababx$]
    Consider the unbordered pattern $w = ab$, which is maximal within the RHS. Then $\ParikhMax{w}{xaxaabbby} \leadsto 2\ParikhVar{w}{x} + 2 + \ParikhVar{w}{y}$ and $\ParikhMin{w}{xyabababx}$ rewrites to $ \ParikhVar{w}{y} + 3 + 2\ParikhVar{w}{x}$. 
    Thus, $\ParikhMax{w}{xaxaabbby} - \ParikhMin{w}{xyabababx}$ is $-1$; witnessing that the equation is unsatisifable.
\end{example}

\section{Implementation and Experiments}
\label{sec:experiments}
\emph{Implementation.} We implemented our approach within a new prototype \textsc{ZIPT}. Our implementation\footnote{available at \url{https://github.com/CEisenhofer/ZIPT}} uses the \textsc{Z3}~\cite{Z3} SMT solver both as the auxiliary integer solver, as well as for the general CDCL($\mathcal{T}$) framework, calling our string solving procedure with the set of plain string equations using user-propagation~\cite{UserPropagator}.

Our implementation follows the workflow of Figure~\ref{fig:overview}, by using the Nielsen transformation rules of Table~\ref{tab:rules} in combination with equality decomposition~\ref{sec:eqSplit}, power introduction~\ref{sec:pwIntr}, and Parikh images~\ref{sec:parikh}. If multiple generating rules of Table~\ref{tab:rules} are applicable, we choose the one predicted to cause faster termination. For example, rules eliminating string variables are prioritised over those that do not.

In addition to the generating rules presented, we use look-ahead heuristics to prefer generating rules in which all but one generated successor node results in a conflict. Examples for such cases are:
\begin{itemize}
    \item Eliminating some string variable $x$ using $x / \varepsilon$ with $xu_1v_1 \simeq u_2v_2$, where $u_1 \in (\{ x \} \cup \VarC \cup \Sigma)^+$ and $u_2 \in (\VarC \cup \Sigma)^+$. Unless a power introduction rule is applicable, we can compute the longest prefix $w$ of $u_2$ such that we can safely apply $x / wx'$. For example,
    $xxbau \simeq abv$ gives $x / abx'$ without branching, as both $x / \varepsilon$ and $x / a$ would yield a conflict immediately. A similar case applies if $x$ was a power token;
    \item Unwinding a power token $w_1^m u \simeq w_2v$ with $w_2 \in (\VarC \cup \Sigma)^+$. Whenever $w_2$ cannot be a prefix or a suffix of any word we can get by unwinding $w_1^m$ at least once, we conclude $m = 0$. For example, $(ab^{m_2}c)^{m_1}u = aav$ enforces $m_1 = 0$ as $aa$ is not a prefix of $ab^{m_2}c$ for any $m_2$ nor is it a suffix;
    \item Length constraints where we can deduce $n \models \Len{x} > \Len{y}$ in our current Nielsen node $n$. If we consider $xu \simeq yv$, we have to apply $x / yx'$. Similarly, if $n \models \Len{x} = \Len{y}$, we chose $x / y$.
\end{itemize}
For traversing and expanding the Nielsen graph, we rely on a variant of iterative deepening, in which successors of node $n$ that strictly reduce the constraints of $n$ are expanded in more depth.\smallskip 

\noindent\emph{Experimental setup.}
As such, we ran our solver \textsc{ZIPT} on the four \verb|QF_S| tracks of the {\tt woorpje} benchmark set~\cite{woorpje} of SMT-LIB~\cite{SMTLIB}; this benchmark set consists of $409$ benchmark files containing only string equations. 
We compared our work against the state-of-the-art solvers competing in the most recent SMT-COMP competition\footnote{~\url{https://smt-comp.github.io/2024/}}. We used a $10$ seconds timeout, $8$ GB of RAM, and default solver configurations, based on a dedicated core of an Intel i7-13850HX CPU. \smallskip 

\begin{table}[t]
    \centering
    \begin{tabular}{|l|c|c|c|c|c|c||c|}
    \hline
    & \textsc{ZIPT} & \textsc{Z3} & \textsc{cvc5} & \textsc{OSTRICH}  & \textsc{Z3-Noodler} & \textsc{Z3str3} & Total\\
    \hline
    \verb|track 01| & 200 & 198 & 191 & 198 & 200 & 200 & 200 \\
    \verb|track 02| & 9 & 4 & 1 & 4 & 6 & 6 & 9 \\
    \verb|track 03| & 195 & 176 & 164 & 127 & 190 & 190 & 200 \\
    \verb|track 04| & 200 & 198 & 200 & 198 & 199 & 200 & 200 \\
    \hline
    \end{tabular}\smallskip
    \caption{Number of solved problems using string solving benchmarks from the {\tt woorpje} benchmark set of SMT-LIB, organised in four tracks. The number of overall problems in each track is listed in column~8. We compare our solver {\tt ZIPT} to the related approaches of \textsc{Z3}~\cite{Z3}, \textsc{cvc5}~\cite{cvc5}, \textsc{Ostrich}~\cite{Ostrich}, \textsc{Z3-Noodler}~\cite{Noodler}, and \textsc{Z3str3}~\cite{z3Str3}.}
    \label{tab:exp}
\end{table}

\noindent\emph{Experimental analysis.} 
Our results are summarized in Table~\ref{tab:exp}, showcasing that {\tt ZIPT} outperforms the state-of-the-art.
When solving problems within  \verb|track 02|, our approach benefits from power introduction; this is so because models described by these benchmarks are exponential in the number of string variables. Using (nested) power tokens allows us to solve such examples without taking exponentially many steps, which is a key difference compared to other works. 

\section{Related Work}
\emph{Decomposing string equations into subsequences} is typically used mainly as a preprocessing step in string solvers. Even though it is primarily only applied in cases where no padding is required, a different variation of our padding has been described in detail in previous works about string solving~\cite{StrDecomposition}. 

\noindent\emph{Recognising self-dependencies in string variables}, especially via explicit powers, is studied in~\cite{StrDecidability,PowerExpr}. While the languages definable by such equations are generally EDT0L~\cite{EDT0L}, the explicit powers we consider fall within more tractable subclasses. Theoretical results about word equations expressing certain power terms date back even longer~\cite{lyndon1962equation}.
In terms of solver support, the approach of~\cite{cvc4StringSolving} can handle some self-dependencies using regular expressions. In contrast, we identify a broader class of self-dependencies and introduce explicit powers, enabling nesting powers for efficiency (see \verb|track 02| in Section~\ref{sec:experiments}).

\noindent\emph{Parikh images}  improve string reasoning via regular expression membership~\cite{Ostrich,ParikhRegex}. Most techniques use single-character Parikh images to detect trivial unsatisfiability. We generalize this by considering Parikh information over multiple character sequences simultaneously. 
Related notions such as Parikh matrices~\cite{ParikhMatrices} and their role in decidability have been explored in~\cite{ParikhUndecidability}. Our approach of analysing potential crossing occurrences is strongly related to recompression~\cite{Recompression}, where such occurrences are stepwise eliminated using case distinctions.

\section{Conclusions}
We introduced a string solving approach, based on Nielsen transformation rules extended with equality decomposition, power introduction, and Parikh images. Our initial results show the practical potential of our work. 
As further work, we plan to improve the analysis via Parikh images by using tighter error approximations and relating the Parikh information from multiple equations rather than considering each equation in isolation. Other steps include heuristically splitting equations using auxiliary variables based on dependency analysis and the introduction of non-ground power terms in order to express dependencies that cannot be represented in our calculus currently.
Further, supporting other string-related SMT-LIB functions, including regular expressions, as defined in the SMT-LIB standard~\cite{SMTLIB}, is an essential next step.


\subsubsection{Acknowledgements.} This research was funded in whole or in part by the  ERC Consolidator Grant ARTIST 101002685, the ERC Proof of Concept Grant LEARN 101213411, the TU Wien Doctoral College SecInt, the FWF SpyCoDe Grant 10.55776/F85,  the WWTF grant ForSmart   10.47379/ICT22007, and the Amazon Research Award 2023 QuAT.

\subsubsection{Disclosure of Interests.}
The authors have no competing interests to declare that are relevant to the content of this article.

\bibliographystyle{splncs04.bst}
\bibliography{refs}

\end{document}